\documentclass[11pt]{article}

\usepackage{graphicx}

\newcommand{\ket}[1]{|#1\rangle}

\oddsidemargin=\evensidemargin
\begin{document}

\title{Vibrational inelastic scattering effects in molecular electronics}

\author{H Ness $^*$ and A J Fisher $^\#$\\
$^*$ CEA-Saclay, Service de Physique et de Chimie des Surfaces\\
et Interfaces, DSM/DRECAM, B\^at 462, 91191 Gif sur Yvette,\\
France, {\tt ness@dsm-mail.saclay.cea.fr};\\
$^\#$ Department of Physics and Astronomy and\\
London Centre for Nanotechnology,\\ 
University College London, Gower Street,\\
London WC1E 6BT, UK, {\tt andrew.fisher@ucl.ac.uk}}

\maketitle

\abstract{
We describe how to treat the interaction of travelling electrons with 
localised vibrational modes in nanojunctions. We present a multichannel 
scattering technique which can be applied to calculate the transport 
properties for realistic systems, and show how it is related to other 
methods that are useful in particular cases. We apply our technique to 
describe recent experiments on the conductance through molecular junctions.}
\endabstract{.}

\section*{Introduction}

Electronic transport through nanoscale systems connected to external
electrodes exhibits a number of new features when compared to
conduction through macroscopic systems. Primary among these is the
importance of local interactions--including Coulomb interactions
between the electrons, and scattering from localised atomic
vibrations. Crudely speaking these effects are more important in
nanoscale systems because the electronic probability density is
concentrated in a small region of space; normal screening mechanisms
are then ineffective.

Perhaps the most extreme instance of nanoscale transport is where an 
electrical connection is formed through a single molecule. This paper 
describes how to approach electron transport in such systems theoretically.  
Our aim is to convince the reader that in many cases the local interaction 
is not a small perturbation which can be added in later---instead, it 
plays the principal role in determining the transport physics.  
In order to treat it, we need to understand the role of the local molecular 
properties---dominated by these local interactions---in scattering 
current-carrying electrons as they pass by. As well as determining the 
current flow, the interactions are then responsible for other important 
phenomena such as local heating.

Such a situation is already familiar in the physics of transport through 
semiconductor quantum dots \cite{kouwenhoven01}, where local Coulomb 
interactions produce Coulomb blockade (regions of very low conductance 
where changes in the charge state of the dot are energetically forbidden) 
and the Kondo effect (where the local spin of the dot electrons is 
screened by the conducting carriers). But the situation in molecules, 
where the length scales are smaller and the electronic energy scales are 
correspondingly larger, is only now being studied.  

If the (three-dimensional) extent of an electronic wavepacket is $L$, the 
Coulomb interaction between electrons scales as $L^{-1}$, while the energy 
arising from short-range interactions with lattice vibrations scales 
as $L^{-3}$ \cite{emin76}.  We might therefore expect the electron-lattice 
interactions to dominate for single molecules where $L$ is small, and 
indeed their signatures have been observed in several recent experiments 
with single molecules (or small numbers) 
\cite{WHo02,HPark00,JKushmerick04,WWang04}.
These experiments include inelastic electron tunneling spectroscopy (IETS) 
of small molecules adsorbed on surfaces \cite{WHo02},
studies of molecular-scale transistors \cite{HPark00}, and experiments on 
molecular junctions
made of alkyl or $\pi$-conjugated molecular wires \cite{JKushmerick04} 
or alkanedithiol self-assembled monolayers \cite{WWang04}.
Such experiments show the possibility of exciting specific vibrational 
modes by injecting electrons; these excitations show up as sharp features 
in the differential conductance.

It has been known for many years that coupling between electrons and molecular 
vibrations plays an important role in  transport properties of bulk films 
of long conjugated molecules (such as conducting and semiconducting polymers);
the transport then occurs by classical diffusion of polarons or 
solitons \cite{AHeeger01}. 
The new realisation is that such effects are also crucial in nanoscale 
transport through individual molecules, even although a classical diffusive 
model for the resulting composite excitations is no longer appropriate.
This new realisation has coincided with an appreciation of the importance of 
local heating \cite{TTodorov98}, another manifestation of the coupling 
between electrons and phonons.

Corresponding to this experimental progress, theorists have
increasingly been studying electron transport in the presence of local
electron-phonon ($e$-ph) coupling in the molecule. In cases where the
$e$-ph effects are weak (for example, in IETS of small molecules,
where the electron residence time on the molecule is very short),
perturbation theory for the $e$-ph interaction can be used to
interpret IETS spectra for model systems \cite{BPersson87} and for
more realistic systems including small molecules on surfaces
\cite{NLorente00} and molecular wires \cite{DiVentra04}. 
Going beyond perturbative approximations, the effects of $e$-ph coupling
have been considered using scattering theory and Green's functions approaches
\cite{JBonca95,HNess99,KHaule99,NMingo00,HNess01,HNess02, ATroisi03},
using master and kinetic equations \cite{PHanggi04}, and reduced
density matrix approaches \cite{ANitzan01}. Another contribution
including dissipation with a semiclassical treatment of the vibration
was also developped \cite{CJoachim02}. More recently, approaches
based on non equilibrium statistical physics have been developped for
model systems \cite{TMii03,MGalperin04} and realistic atomic and
molecular wires \cite{APecchia04,YAsai04,TFrederiksen04}.

In this paper, we describe the effects of the coupling between
injected electrons and molecular vibrations (both treated on the
quantum level) by using a multi-channel inelastic scattering
technique. We show that it is important to treat such local
interactions non-perturbatively, and how our approach relates
to others. In the final part we study the effects of the
temperature on the conductance properties and shed some light on the
basic understanding of the features observed in IETS spectra.

Note that in the following, we consider independent charge carriers 
interacting with single or multiple quantized vibrational modes. 
Calculations including the interaction between charges 
\cite{PCornaglia04} or the interaction between charge and classical
vibrations \cite{AHorsfield04} have been considered elsewhere.

\section*{Model}

Before presenting some numerical results, we describe our approach to
inelastic transport and its relation to other approaches.

\subsection*{Physical model}

Our model involves those delocalized electron states which
simultaneously carry current through the molecule and interact with
its eigenmodes of vibration. In the case of conjugated molecules
these states are predominantly derived from the $\pi$ orbitals (though
our method can deal equally with other types of state,
and there are no restrictions on the geometry or dimensionality).
If the molecular vibrations are approximately harmonic, the 
reference Hamiltonian $H_0$ for the molecule is
\begin{equation}\label{eq:basicH}
H_0=H_{\rm el}
+\sum_\lambda\omega_\lambda a^\dag_\lambda a_\lambda
\label{H0}
\end{equation}
where $H_{\rm el}$ is the purely electronic Hamiltonian and $\lambda$
labels the eigenmodes of vibration of the isolated molecule in its
equilibrium (generally neutral) charge state; $a^\dag_\lambda$ creates
($a_\lambda$ annihilates) a quantum of energy $\omega_\lambda$ in mode
$\lambda$. Note that $H_{\rm el}$ may, in principle, contain
electron-electron interactions.

Our fundamental approximation is to consider the scattering only of
\textit{single} electrons or holes from the molecule. We expect this
assumption to be reasonable when the interval between carrier
transmission events is much greater than the transit time through the
molecule \cite{HNess01}, and provided that the applied bias does not
cause significant fluctuations in the occupancy of the molecular
eigenstates (i.e. provided one charge state of the molecule dominates
during the transport process). 

For electron transport, if the
equilibrium charge state contains $N$ electrons and $\ket{0,N}$ is the
$N$-electron ground state, we construct an electronic basis
$\ket{i,N+1}=c^\dag_i\ket{0,N}$ where the \{$c^\dag_i$\} create an
electron in a sufficiently complete set of orbitals (for example, in
all the low-lying $\pi$-states of the molecule). For hole transport
we would use $\ket{j,N-1}=c_j\ket{0,N}$. We now diagonalize $H_{\rm el}$
(keeping the phonon coordinates fixed, for the moment) within
our restricted $(N\pm1)$-electron basis to obtain a set of approximate
$(N\pm1)$-electron eigenstates. Using the electron creation
($c^\dag_n$) and annihilation ($c_n$) operators and the energies
$\epsilon_n$, we can write
\begin{equation}
H_{\rm el}=\sum_n\epsilon_nc^\dag_nc_n.
\end{equation}
The approximate $(N\pm1)$-electron eigenstates of the full $H_0$ can thus be 
written as $\ket{n,\{n_\lambda\}}$, where $n$ labels an electronic eigenstate 
and the $\{n_\lambda\}$ are occupation numbers for the vibrational modes:
\begin{equation}
\ket{n,\{n_\lambda\}}=c^\dag_n\prod_\lambda 
(a^\dag_\lambda)^{n_\lambda}/\sqrt{n_\lambda!}\ \vert 0,N,\{0_\lambda\}\rangle
\end{equation}
and $\vert 0,N,\{0_\lambda\}\rangle$ is the electronic and vibrational 
ground state for $N$ electrons.

The {\it e}-ph coupling term $H_{\rm {\it e}ph}$ is taken to be linear in the 
phonon displacements and induces transitions between these electronic states:
\begin{equation}
H_{\rm {\it e}ph}=
\sum_{\lambda,n,m}\gamma_{\lambda nm}(a^\dag_\lambda+a_\lambda)
c^\dag_n c_m\ .
\label{Heph}
\end{equation}
The values of $\epsilon_n$, $\omega_\lambda$ and 
$\gamma_{\lambda nm}$ are 
calculated from a suitable model of the isolated molecule.  
In our work on conjugated molecules \cite{HNess01} we have used the 
Su-Schrieffer-Heeger (SSH) \cite{AHeeger01} model within the harmonic limit.

To obtain the transport properties, we perform a thought experiment in which we
connect the  left and right ends of the molecule to metallic leads and 
scatter a single incoming charge carrier (electron or hole).
The coupling matrix elements are $v_{L,R}$ respectively; since we do not wish 
to focus on the properties of the leads we take the simplest possible model 
for them and consider one-dimensional semi-infinite chains (with on-site 
energy $\epsilon_{L,R}$ and hopping integrals $\beta_{L,R}$, giving dispersion 
relations $\epsilon=\epsilon_{L,R}+2\beta_{L,R}\cos(k_{L,R})$, where $k_{L,R}$ 
are dimensionless wavevectors).  
We assume transport is purely elastic within the leads themselves; 
dissipation  occurs in remote reservoirs,  as in the standard Landauer 
picture for electron transport.
The scattering states $\vert\Psi\rangle$ for a single incoming carrier 
are then expanded inside the molecule onto the eigenstates
$\vert n,\{n_\lambda\}\rangle$ of $H_0$.
The single added carrier can be anywhere in the system and interacts with
lattice vibrations only when inside the molecule.  

The transport problem is solved by mapping the many 
body problem onto a single-electron one 
with many scattering channels \cite{JBonca95,KHaule99,HNess01}; 
each channel represents a process by which 
the electron might exchange energy with the vibration modes.  
For an initial mode distribution $b\equiv\{m_\lambda\}$ and 
an incoming electron from the left, the outgoing channels in the left 
and right leads are associated with energy-dependent reflection 
coefficients $r_{ab}(\epsilon)$ and transmission coefficients 
$t_{ab}(\epsilon)$, where $a\equiv\{n_\lambda\}$ is the final mode 
distribution.
In the leads, the scattering states are propagating
waves with amplitudes $r_{ab}$ (reflection) and $t_{ab}$ (transmission),
and wave vectors corresponding to the initial ($\epsilon_{\rm in}$) and 
final ($\epsilon_{\rm fin}$) electronic energies.
By projecting out the leads, we work entirely in the molecular subspace 
to obtain the scattering state $\vert\Psi\rangle$ by solving 
\begin{eqnarray}
\left[
\omega-H_0-H_{e{\rm ph}}-\Sigma^r_L(\omega)-\Sigma^r_R(\omega)
\right]\
\vert\Psi(\omega)\rangle \nonumber \\ 
\equiv G^r(\omega)^{-1}\ket{\Psi(\omega)}=\vert s_b(\omega)\rangle.
\label{linsys}
\end{eqnarray}
Here $\omega$ is the conserved total energy; 
\begin{equation}\label{eq:econserved}
\omega=\epsilon_{\rm in}+\sum_\lambda m_\lambda\omega_\lambda
=\epsilon_{\rm fin}+\sum_\lambda n_\lambda\omega_\lambda.
\end{equation}
The source term $\vert s_b(\omega)\rangle$ is fixed by the incoming 
boundary conditions, while $\Sigma^r_{L,R}(\omega)$ are the electron 
self-energies arising from the coupling of the molecule to the leads. 
The components of the scattering states 
then give the matrix elements of the retarded Green's function $G^r(\omega)$. 
In practice, the linear system $\vert\Psi\rangle=G^r\vert s\rangle$ 
is solved for a finite-size molecular subspace, 
by truncating the occupation number above a maximum 
$n_{\lambda}^{\rm max}$ in each mode.
This is physically reasonable because the injected
charge cannot populate infinitely many excitations.

From the solution of Eq.(\ref{linsys}), we can calculate any observable 
property, such as the expectation value of an operator $A$:
$\langle A\rangle = \sum_{\{b\}} \mathcal{W}_{b}^{\rm ph}(T)
\langle\Psi\vert A \vert\Psi\rangle$
where $\mathcal{W}_{b}^{\rm ph}(T)$ is the statistical weight
of the distribution $b\equiv\{m_\lambda\}$ of 
modes at temperature $T$.  
The transmission probability $T_{ab}$ is given from the square of
the transmission coefficients $t_{ab}$ with the usual ratio of the electron
velocities in the outgoing and incoming channels.
In our model,
$t_{ab}\propto\langle i=N\vert G_{ab}\vert s\rangle$
and the transmission probability $T_{ab}$ is written 
as \cite{HNess02,ATroisi03}
\begin{eqnarray}
T_{ab}(\epsilon_{\rm fin},\epsilon_{\rm in})
=4\ \frac{v_L^2}{\beta_L}\sin k_b^L(\epsilon_{\rm in})\  
  \frac{v_R^2}{\beta_R}\sin k_a^R(\epsilon_{\rm fin})\ \nonumber \\
  \times \vert\langle i=N\vert G^r_{ab}(\omega)\vert i=1\rangle\vert^2 \ ,
\label{Tinel}
\end{eqnarray}
where 
$\langle N\vert G^r_{ab}(\omega)\vert 1\rangle$ is the matrix element of
the Green's function $G^r$ taken between the left side $i=1$ and the right 
side $i=N$ of the molecule and the vibration distributions before ($b$) 
and after ($a$) scattering.
The factors ${v_{L,R}^2}/{\beta_{L,R}}\sin k_{b,a}^{L,R}$ are 
related to the imaginary parts of the retarded self-energies 
$\Sigma^r_{L,R}$.

Two concluding notes are in order.  First, this procedure solves 
the problem \textit{non-perturbatively} in the $e$-ph interaction.
Second, we have described the method for the case where the channel structure 
of the leads is generated only by the vibrational excitations of the molecule, 
but it is straightforward to generalise it to the case of multiple spatial 
channels in the leads. 

\subsection*{Connection to other methods}

Here we etablish how the multichannel scattering technique 
is related to other methods based on
two-particle Green's functions for non-interacting electrons
\cite{NWingreen89} and on non-equilibrium Green's functions 
\cite{MGalperin04,APecchia04,YAsai04,TFrederiksen04}.
We will show below (for a simplified molecular model)
that $G^r$ can be reformulated to include the self-energies arising from 
the coupling of the molecule to the leads ($\Sigma^r_{\rm leads}$) 
and from the interaction of the electron with the vibrations 
($\Sigma^r_{e{\rm ph}}$).

Let us start with the generalisation of the Landauer formula
for non-equilibrium interacting systems given in Ref.\cite{YMeir92}. 
In the following we consider the single-site, single-vibrational mode 
(SSSM) model, which allows us to illustrate the essential physics without 
complicating the algebra.
We first assume that the left and right leads are identical, 
and couple to the molecule via energy-independent hopping integrals. 
Then, the self-energies $\Sigma^r_{L,R}$ (and their imaginary
parts $\Gamma_{L,R}$) are proportional to each other. 
According to Ref.\cite{YMeir92}, the current through the junction is 
then given by
\begin{eqnarray}
I=\frac{2e}{h} 
\int d\omega \left( f_L(\omega)-f_R(\omega)\right)
{\rm Im Tr}\{\Gamma(\omega)G^r(\omega)\} \ ,
\label{Imeir}
\end{eqnarray}
where $f_{L,R}$ is the Fermi distribution of the left and right
(non-interacting) lead respectively,
$\Gamma\equiv\Gamma_L\Gamma_R/(\Gamma_L+\Gamma_R)$ and
$G^r$ is the retarded Green's function of the molecule including the 
self-energies arising from the coupling to the leads and from the 
interaction between particles.
Here we consider non-interacting charge carriers coupled to vibrations
of the molecule.

Within the SSSM model there is only one energy level $\epsilon_0$
that couples to a single mode of frequency $\omega_0$ with a coupling 
constant $\gamma_0$.
For this model, $G^r$ is the inverse of a tridiagonal matrix $R$ in
the subspace $\vert\chi_n\rangle$ of the excitations $n$ of the 
vibrational mode ($n=0$ is ground state of the mode) with 
diagonal elements 
$R_{n,n}=\omega-\epsilon_0-n\omega_0-\Sigma^r_{\{n\}}(\omega)$ and
non-zero off-diagonal elements 
$R_{n,n+1}=-\gamma_0\sqrt{n+1}$ and $R_{n,n-1}=-\gamma_0\sqrt{n}$.
Here $\Sigma^r_{\{n\}}(\omega)$ is the retarded self-energy of 
both leads for the channel containing $n$ phonon excitations; it is 
calculated by taking into account the energy conservation condition 
Eq.(\ref{eq:econserved}), which reduces in the limit of very 
low temperature to 
$\omega=\epsilon_{in}=
\epsilon_{fin}+n \omega_0$.
Furthermore, in this limit, only the matrix element 
of $G^r$ between the phonon ground state 
$G^r_{00}=\langle\chi_0\vert G^r(\omega)\vert\chi_0\rangle$ 
enters in the evaluation of the current.
It can be expressed as a continued fraction:
\begin{eqnarray}
G^r_{00}(\omega) =
[
\omega-\epsilon_0-\Sigma^r_{\{0\}}(\omega)- \nonumber \\
\gamma_0^2/(
\omega-\epsilon_0-\omega_0-\Sigma^r_{\{1\}}(\omega)- \nonumber \\
2\ \gamma_0^2/(
\omega-\epsilon_0-2 \omega_0-\Sigma^r_{\{2\}}(\omega)- \nonumber \\
3\ \gamma_0^2/(
\omega-\epsilon_0-3 \omega_0-\Sigma^r_{\{3\}}(\omega)- ...
]^{-1}
\label{Gr_contfrac}
\end{eqnarray}

The imaginary part of $G^r_{00}$ can be rewritten 
as the sum of several terms:
\begin{eqnarray}
{\rm Im} G^r_{00}(\omega) = 
\vert G^r_{00}(\omega)\vert^2\ \times \nonumber \\
{\rm Im}
\left\{
\Sigma_{\{0\}}+
\frac{\gamma_0^2}{\omega-\epsilon_0-\omega_0-\Sigma^r_{\{1\}}- ...}
\right\} \ .
\label{ImGr_A}
\end{eqnarray}
After some calculation, it can be shown that
${\rm Im} G^r_{00}$ is actually related to the other non-diagonal elements
$\langle\chi_n\vert G^r\vert\chi_0\rangle$ of the Green's function 
and that finally one has
\begin{eqnarray}
{\rm Im} \langle\chi_0\vert G^r\vert\chi_0\rangle
= \sum_{n} {\rm Im} \Sigma^r_{\{n\}}\
\vert\langle\chi_n\vert G^r\vert\chi_0\rangle\vert^2 \ .
\label{ImGr_B}
\end{eqnarray}

Introducing this result in Eq.(\ref{Imeir}), the current for the SSSM model 
is then expressed as the sum of the elastic and inelastic 
transmission probabilities $T_{ab}(\epsilon',\epsilon)$ obtained 
from Eq.(\ref{Tinel}).

One can perform a similar derivation at a finite temperature $T$.
The current is then determined from all the diagonal matrix elements
${\rm Im}\ \langle\chi_n\vert G^r\vert\chi_n\rangle$ including the 
appropriate statistical weight $\mathcal{W}_n^{\rm ph}(T)$
for the $n$-th excitation of the vibration mode.
Each matrix element ${\rm Im}\ \langle\chi_n\vert G^r\vert\chi_n\rangle$ 
is related to the sum of the elastic and inelastic transmission coefficients
where the vibration mode is in the $n$-th excited state before scattering,
and in the $m$-th excited state afterwards. 

The derivations given above can be extended to general cases in which 
many electronic levels are coupled to many vibration modes.
Then the $R$ matrix which defines $G^r$ is block-tridiagonal; the 
diagonal blocks correspond to purely electronic processes with constant 
phonon occupancy, while the off-diagonal blocks correspond to
the $e$-ph coupling matrix $\gamma_{\lambda nm}$.
The equivalent matrix element
$\langle\{0_\lambda\}\vert G^r\vert\{0_\lambda\}\rangle$ 
of Eq.(\ref{Gr_contfrac}) is then written as a matrix continued fraction 
in the corresponding basis. 
The same reasoning derived above holds for calculating the current in terms 
elastic and inelastic transmission probabilities $T_{ab}$ for all temperatures.

Now, we show how to connect Eq.(\ref{ImGr_B}) to the approach
developed in \cite{NWingreen89}.
For this, let us consider the wide-band limit for the leads,
so the lead self-energies become independent
of the energy and hence of the channel considered, and are purely 
imaginary 
$\Sigma^r_{n}(\omega)\equiv\ i\Gamma=i(\Gamma_L+\Gamma_R)$. 
The imaginary part of $G^r_{00}$ then becomes
${\rm Im}\ G^r_{00}(\omega)=\Gamma\sum_n 
\vert\langle\chi_n\vert G^r\vert\chi_0\rangle\vert^2$.
It is convenient to work in this limit to circumvent
the problems of band-width renormalisation when
one introduces the transformation $U$ that diagonalises
$H=H_0+H_{e{\rm ph}}=\epsilon_0 c^\dag_0 c_0+\omega_0 a^\dag_0 a_0
+\gamma_0(a^\dag_0+a_0) c^\dag_0 c_0$.
We introduce the basis set 
$\vert\tilde{\chi}_n\rangle=U\vert\chi_n\rangle$,
and obtain
\begin{eqnarray}
{\rm Im}\ G^r_{00}= \Gamma \sum_n
\left|\sum_l
\frac{\langle\chi_n\vert\tilde{\chi}_l\rangle
\langle\tilde{\chi}_l\vert\chi_0\rangle}
{\omega-\tilde\epsilon_0-l \omega_0 + i \Gamma}\
\right|^2 \ ,
\label{ImGr_WBL_1}
\end{eqnarray}
where $\tilde\epsilon_0=\epsilon_0-g^2\omega_0$ and
$g=\gamma_0/\omega_0$.
Then, calculating the overlaps $\langle\chi_n\vert\tilde{\chi}_l\rangle$
between the original $\vert\chi_n\rangle$ and displaced harmonic states 
$\vert\tilde{\chi}_n\rangle$ as in \cite{Glauber69}, one ultimately finds
\begin{eqnarray}
{\rm Im}\ G^r_{00}(\omega)= \Gamma e^{-2g^2}\ \sum_{n=0}^{\infty} 
\frac{g^{2n}}{n!} \times \nonumber \\
\left|\sum_{j=0}^{n}(-1)^j {n \choose j}
\sum_{l=0}^{\infty}
\frac{g^{2l}}{l!}
\frac{1}
{\omega-\tilde\epsilon_0-(j+l) \omega_0 + i \Gamma}\
\right|^2 \nonumber \\
\label{ImGr_WBL_2}
\end{eqnarray}
Eq.(\ref{ImGr_WBL_2}) is just the result derived in \cite{NWingreen89}
for the transmission by starting from a two-particle Green's
function description for a single resonant level coupled to a single
vibration mode.

It should be noticed that the equivalence between the two approaches
is only valid in the wide-band limit for which the full polaron shift
is obtained (i.e., $\epsilon_0$ shifted by the relaxation energy
$-\gamma_0^2/\omega_0$).  In other cases, the energy dependence of the
self-energies $\Sigma^r_{\{n\}}$ plays an important role.  Furthermore
when the residence time of the electron in the molecule is not long
enough for the vibration to respond fully to its presence, the full
relaxation is not obtained and one has an intermediate polaron
shift\footnote{It should be noted that no polaron shift is obtained 
if one makes a perturbative expansion of the $e$-ph coupling.}.
However there are no such limitations in the multichannel scattering
technique, which can treat the problem for the full parameter range
(including strong $e$-ph coupling)---provided only that the fundamental 
assumption of single-carrier transport (see the discussion
following Eq.(\ref{eq:basicH})) remains valid.

Finally, in this section, we show how the multichannel scattering technique
is related to more recent approaches based on non-equilibrium Green's 
functions \cite{TMii03,MGalperin04,APecchia04,YAsai04,TFrederiksen04}.
Such a non-equilibrium approach is more suitable when our assumption of a 
well-defined reference charge state breaks down.
Let us first rewrite the retarded Green's function in Eq.(\ref{Gr_contfrac}) as
\begin{equation} 
G^r(\omega)=[\omega-\epsilon_0-\Sigma^r_{\{0\}}(\omega)
-\Sigma^r_{e{\rm ph}}(\omega)]^{-1} \ , 
\label{otherGr}
\end{equation}
where the self-energy due to the interaction between electron
and phonon is {\it symbolically} and {\it recursively} given by
$\Sigma^r_{e{\rm ph}}(\omega)\equiv \gamma_0^2\ G^r(\omega-\omega_0)$
\footnote{This is a closed form for a lowest order series expansion of the 
$e$-ph self-energy equivalent to the self-consistent Born 
approximation \cite{MGalperin04,APecchia04,YAsai04,TFrederiksen04}.
In fact the continued fraction expansion of $\Sigma^r_{e{\rm ph}}$
includes a factor $n$ at each level of the fraction as seen in 
Eq.(\ref{Gr_contfrac}). Such factors arise from applying the creation 
(annihilation) phonon operator on the vibrational state 
$\vert\chi_{n-1(n)}\rangle$ in a multi-excitation process.
The exact series expansion for $\Sigma^r_{e{\rm ph}}$ includes higher-order 
terms corresponding to multiple excitations, for example: 
$\gamma_0^4\ G^r(\omega-\omega_0)G^r(\omega-2\omega_0)G^r(\omega-\omega_0)$.
A similar result can be obtained by a linked cluster expansion approach.}. 
Note that by definition \cite{HNess01}, one has the
following relation for the lead self-energy:
$\Sigma^r_{\{0\}}(\omega-n\omega_0)=\Sigma^r_{\{n\}}(\omega)$.

In a many-body non-equilibrium Green's functions approach,
the self-energies arising from the interaction between particles
are obtained from a diagrammatic perturbation expansion of the
interaction and by applying the rules for the time ordering and for 
the evaluation of products of double-timed operators on the Keldysh 
contour.  
In principle, for a coupled $e$-ph system, one should include processes 
to all orders in the interaction to calculate self-consistently the 
different electron Green's functions dressed by the phonons, as well as 
the phonon Green's functions dressed by the electrons.
This is a tremendous task to achieve for realistic systems, and so far
it has only been done almost exactly for model systems or within some 
approximations for atomic or molecular wires.
Recent studies have been performed using the so-called self-consistent 
Born approximation \cite{MGalperin04,APecchia04,YAsai04,TFrederiksen04},
in which the retarded electron 
self-energy due to $e$-ph coupling is obtained from the sum of several 
contributions all written in the following form:
$\Sigma^{r}_{e{\rm ph},XY}(\omega)\propto
i\gamma_0^2\int d\omega' D^{X}(\omega-\omega')G^{Y}(\omega')$, where the
superscripts $X,Y\equiv r,>$ represents the different types 
of Green's functions
(retarded $r$, Keldysh greater $>$) for the electrons ($G$) and for the phonons ($D$).

In the following, we work with the undressed phonon Green's functions 
$D_0(\omega)$ for which the Keldysh greater component is  
$D^>_0(\omega)=-i2\pi (N(\omega)\delta(\omega+\omega_0)
+(N(\omega)+1)\delta(\omega-\omega_0))$,
$N(\omega)$ being the Bose-Einstein distribution function.
In the limit of low temperatures, simplifications arise because
then $N(\omega)=0$.
Furthermore by using the principle of causality and the analytic properties 
of $G^r$ in the complex plane, it can be shown that the contribution 
$\Sigma^{r}_{e{\rm ph},rr}(\omega)
=-i\gamma_0^2\int d\omega'/2\pi\ D^r_0(\omega-\omega')G^r(\omega')$ 
vanishes\footnote{This is always true for electron-hole-symmetric systems. 
It is also true when the system is not too far from equilibrium.}.
The second contribution to the self-energy is given by
$i\gamma_0^2\int d\omega'/2\pi\ D^>_0(\omega-\omega')G^r(\omega')$. 
At zero temperature, this is exactly the self-energy 
$\gamma_0^2\ G^r(\omega-\omega_0)$ derived previously in Eq.(\ref{otherGr}).
The final contribution to the self-energy is
$\Sigma^{r}_{e{\rm ph},r>}(\omega)\propto
\int d\omega' (D^r_0(\omega')-D^r_0(\omega=0)) 
G^>(\omega-\omega')$.
It involves the greater electron Green's function $G^>$ which provides 
information about the non-equilibirum density of unoccupied states of 
the molecule.
$G^>$ is related to the corresponding self-energy 
$\Sigma^>=\Sigma^>_{\rm leads}+\Sigma^>_{e{\rm ph}}$ and
the retarded and advanced Green's functions $G^{r,a}$ via the
kinetic equation $G^>=G^r\Sigma^>G^a$.
The terms involving $\Sigma^>_{e{\rm ph}}$ are intrinsically of 
higher order in $\mathcal{O}(\gamma_0^2)$ than 
$\Sigma^r_{e{\rm ph}}(\omega)=\gamma_0^2\ G^r(\omega-\omega_0)$ and
can be neglected for weak $e$-ph coupling.
The other terms involving $\Sigma^>_{\rm leads}$ are important;
however their contributions become small when the electronic level coupled 
to the vibration is off-resonance with the Fermi levels of the leads (in 
the limit of zero to small applied bias).
This is case for molecular wires having a substantial 
HOMO-LUMO gap, when the Fermi levels are pinned inside this gap.

So, the multichannel scattering technique takes into account the most 
important contributions to the $e$-ph self-energy.
There are some terms arising from the Keldysh approach to transport 
that are not treated in the multichannel technique; however, as mentioned 
above, the general problem of the non-equilibrium transport is so complex 
that it has only been solved for model systems and using approximations 
for the Green's functions and 
self-energies \cite{TMii03,MGalperin04,APecchia04,YAsai04,TFrederiksen04}.

Now that we have established how to relate the multichannel scattering
technique to other widely used non-perturbative approaches for transport 
through $e$-ph coupled systems, we turn to some practical applications 
of the technique.

\section*{Results}

In our previous studies on long conjugated molecules
\cite{HNess02,HNess01}, we have already shown that the
longitudinal optic phonon modes are the most strongly coupled to the 
injected charges.
We have elucidated the mechanisms of charge injection
and transport in such molecular wires. The transport is
associated with the formation and propagation of polarons in
perfectly dimerized (semiconducting) molecular wires \cite{HNess01}. 
More complex mechanisms arise in the presence of mid-gap states 
involving the delocalisation of a soliton
and polaron-soliton interactions \cite{HNess02}.

Here we present new results for models of molecules in relation with 
recent experiments; namely
we consider the effects of the temperature on the conductance peaks 
around the HOMO-LUMO gap for short molecular wires \cite{JReichert03}, 
and we point out the difficulties for understanding the features 
observed in IETS \cite{JKushmerick04,WWang04} in terms of single 
vibration mode analysis.
In doing so, we provide a plausible explanation for the temperature
dependence of the width and shape of conductance peaks from a model 
system.   More importantly, we show that it is crucial to treat the 
coupling to all the relevant vibration modes simultaneously, 
instead of considering one mode at a time, in order to be able to 
identify the origin of the features in IETS spectra.

\subsection*{Temperature effects on the conductance peaks}

Here we consider the experiments performed with a mechanically 
controlled  breakjunction at room temperature and at low 
temperature ($T\approx$ 30K) for polyphenylene-based molecular 
wires \cite{JReichert03}.
The conductance curves show a peak just above a region of very
low conductance at small bias (which might be a HOMO-LUMO gap or 
Coulomb gap).
At low temperature, the peak is asymmetric with a maximum width 
of $\sim$125 meV. 
At room temperature, this peak has a reduced amplitude and is 
more symmetric with a bell-like shape of maximum width 300 meV.
It is clear that such a temperature dependence cannot be
explained by the broadening of the Fermi sea of the electrodes:
a broadening over $kT$ is not enough to explain the widening of 
the conductance peak and its change of shape.
We therefore assume that there are fluctuations within the 
junction or at the molecule-electrode contacts, and that these
fluctuations are related to low frequency vibration modes.
It is difficult to identify exactly the nature of such modes,
so we consider for simplicity a SSSM model in
which the single level $\epsilon_0$ is coupled to a low
frequency vibration mode ($\omega_0\approx$ 80 meV)
via a coupling constant $\gamma_0\approx\omega_0/2$.

\begin{figure}
\begin{center}
\includegraphics[width=12cm]{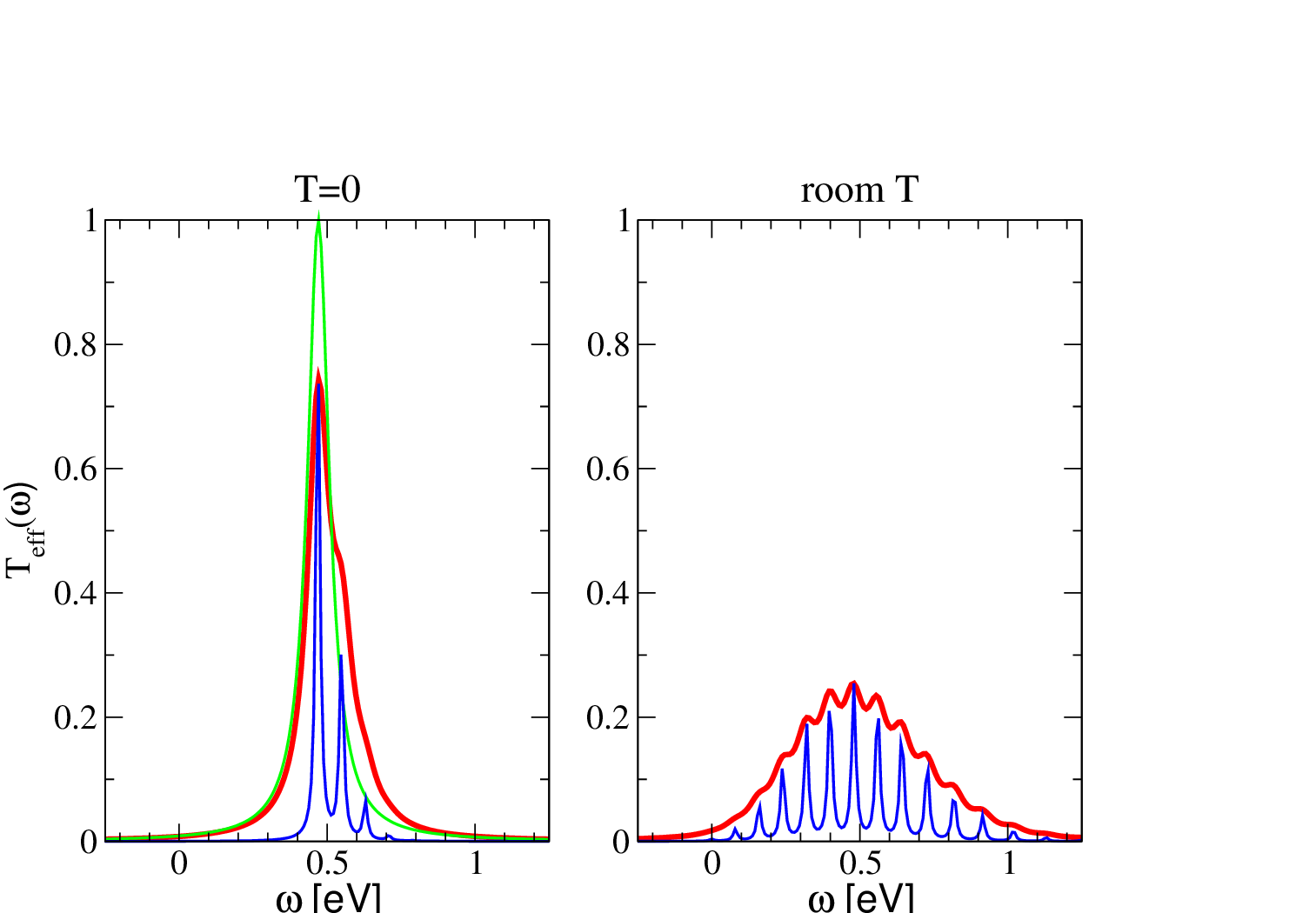}
\end{center}
\caption{Total effective transmission $T_{\rm eff}$ versus energy 
$\omega$
for the SSSM model ($\epsilon_0\approx$ 0.5 eV, $\omega_0\approx$ 80 meV,
$\gamma_0\approx\omega_0/2$).
{\it Left panel}:
Very low temperature limit: the transmission
peak (red line) is asymmetric towards higher energy because of 
the presence of phonon side-bands corresponding to phonon emission.
The resonance peak obtained in the absence of $e$-ph coupling 
(green line) is symmetric around $\epsilon_0$.
{\it Right panel}:
At room temperature, both phonon emission and adsorption processes
are available.
The transmission peak (red line) is more symmetric around $\epsilon_0$ 
and its amplitude is reduced.
In both panels, the blue lines represent the transmission obtained
for a strongly reduced broadening
in order to show explicitly the phonon side-bands.}
\label{fig0}
\end{figure}

We perform calculations for this model according to the 
prescriptions given in the previous section at very low 
temperature and at room temperature.
The total effective transmission $T_{\rm eff}(\omega)$ 
is obtained as the sum of all the elastic and inelastic
contributions to the transmission and is shown in Fig.\ \ref{fig0} 
for the zero temperature limit ($kT=0$) and
for room temperature ($kT\approx\omega_0/2$).
In the absence of $e$-ph coupling, the transmission is given by a 
single symmetric resonance located around $\epsilon_0$, as expected. 
In the presence of $e$-ph coupling, the shape of the resonance
is modified due to the presence of phonon side-bands.
For very low temperatures, the transmission peak is asymmetric towards 
higher energy. 
The phonon side-band peaks are located at energies above $\epsilon_0$
because only phonon emission processes are allowed at low temperatures.
At room temperature, both phonon emission and absorption processes
become possible and the phonon band peaks appear for energies
above and below the molecular level.\footnote{This lineshape appears 
similar to that obtained for very low temperature and very strong 
coupling \cite{Flensberg03}, but the physics is quite different because 
absorption processes are important.}
The transmission peak is then more symmetric around $\epsilon_0$.
Futhermore, its amplitude is reduced compared to the 
peak at $T=0$, as observed experimentally.

It is interesting that such a simple model provides
results in qualitative agreement with the
conductance measurements performed in Ref.\cite{JReichert03}.
Such a mechanism therefore provides a plausible explanation for 
the temperature dependence of the conductance peaks in molecular 
junctions, although we cannot rule out other possible explanations
such as those given in Refs. \cite{TMii03,MGalperin04}.

\subsection*{Multi-vibration mode effects}

We now consider a more realistic molecular wire built from an odd
number $N$ of monomers as obtained from the SSH model.  In this case,
there is a mid-gap state in the HOMO-LUMO gap.  In the following, we
couple all the unoccupied $\pi$ states to the low energy
optic modes as in Eq.(\ref{Heph}).  The actual spatial and frequency
structure of these modes is very important in allowing charge to
propagate through the molecule---in contrast to previous studies on
simpler model systems \cite{JBonca95}.

As a generalisation of the SSSM model, there will also be phonon
side-bands in the transmission.  The phonon side-band peaks related to
the mid-gap state will appear inside the HOMO-LUMO gap.  Such peaks
also exist in the absence of the mid-gap state, i.e. for wires of even
length $N$; however, there they are hidden in the background of the
resonances from the molecular conduction band.

\begin{figure}
\begin{center}
\includegraphics[width=12cm]{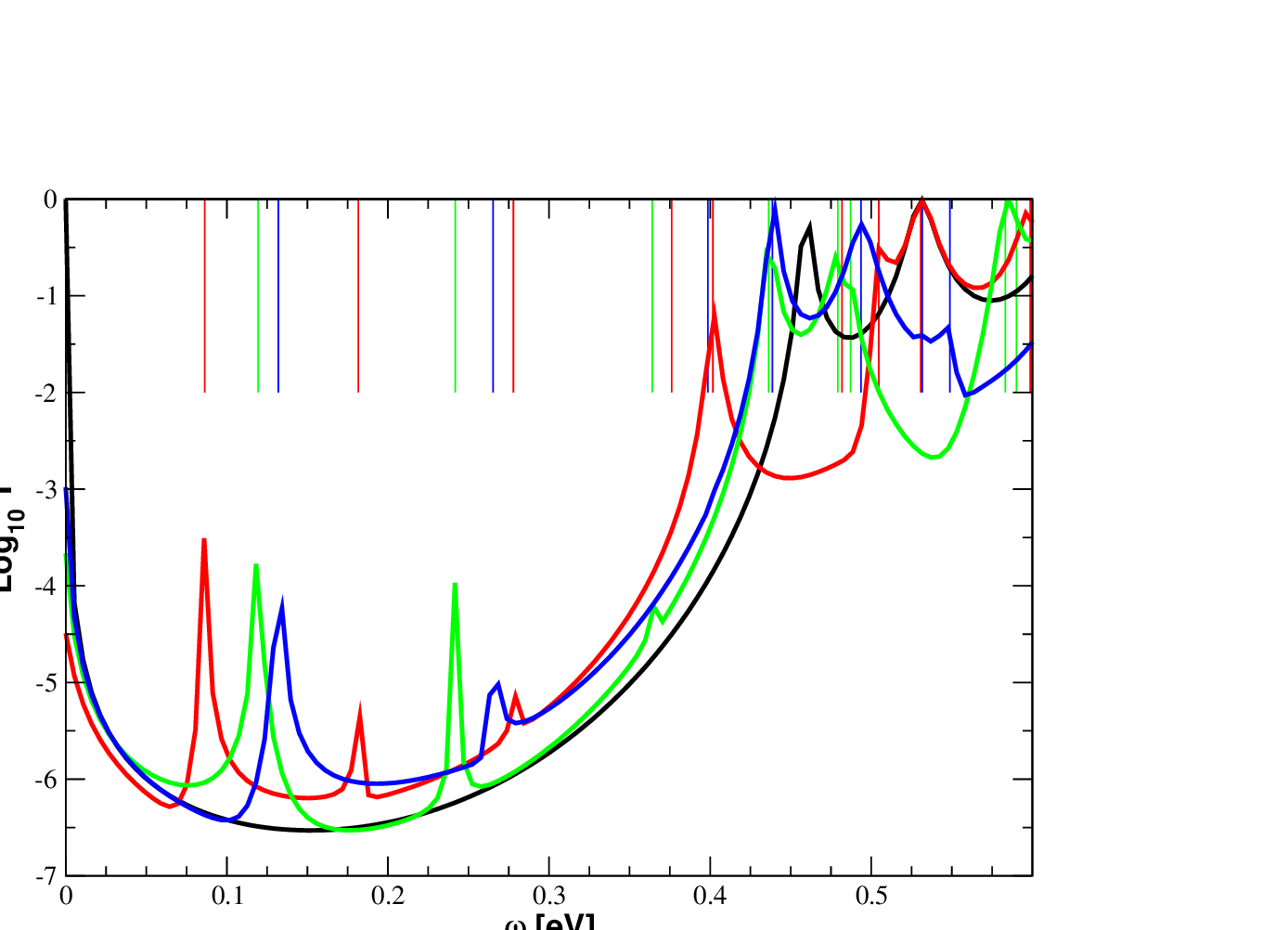}
\end{center}
\caption{Total effective transmission $T_{\rm eff}$ versus 
energy $\omega$ for a molecular wire of length $N=155$. 
The transmission is calculated by including a single vibration 
modes one at a time. 
Optic mode $\omega_{\lambda_1}=112.1$ meV
(red line) showing the most important polaron shift, 
$\omega_{\lambda_2}=129.9$ meV (green line),
$\omega_{\lambda_3}=136.8$ meV (blue line).
The transmission in the absence of $e$-ph coupling is also 
shown (black line).
The vertical bars represent (for the same color) the eigenvalues
of the isolated molecule Hamiltonian including the coupling of 
the single mode $\lambda$ to the many molecular levels.}
\label{fig1}
\end{figure}

The transmission curves for a wire of length $N=155$ in
which the molecular levels interact with a single vibration mode 
one at a time are shown in Fig.\ref{fig1}.
Several different optic modes 
($\omega_{\lambda_1,\lambda_2,\lambda_3}=$ 112.1, 129.9, 136.8 meV) 
are considered. 
The total effective transmission $T_{\rm eff}$ shows phonon side-band 
peaks in the HOMO-LUMO gap,
separated as expected by the corresponding energy $\omega_\lambda$.
Up to small energy shifts due to real part of the self-energies, the 
peaks in the transmission correspond to the eigenvalues of the coupled 
$e$-ph Hamiltonian of the isolated molecule, as shown by the vertical 
bars in Fig.\ref{fig1}.
For some eigenvalues, there are no resonances in the transmission. 
There are several possible reasons for this: 
(i) the phonon side-band peaks associated to a given electronic
transition have an exponentially decaying weight for large numbers
of excitation quanta, so some of these peaks may disappear in the
tail of other resonances; (ii) there might be destructive
quantum interferences for the corresponding process\footnote{Such 
effects appear more often when several electronic levels are coupled 
simultaneously to several vibration modes.}; 
(iii) the corresponding eigenstates are such that the matrix elements 
$\langle N\vert G^r_{ab}\vert 1\rangle$ in Eq.(\ref{Tinel}) are small 
or vanishing. 

Note that such calculations also reveal which optic phonon mode 
contributes the most to the polaron shift (shift towards lower energy 
of the first resonance above the gap due to the $e$-ph coupling).
In the present case, the optic mode of energy
$\omega_{\lambda_1}=$ 112.1 meV gives the most important polaron shift.

\begin{figure}
\begin{center}
\includegraphics[width=12cm]{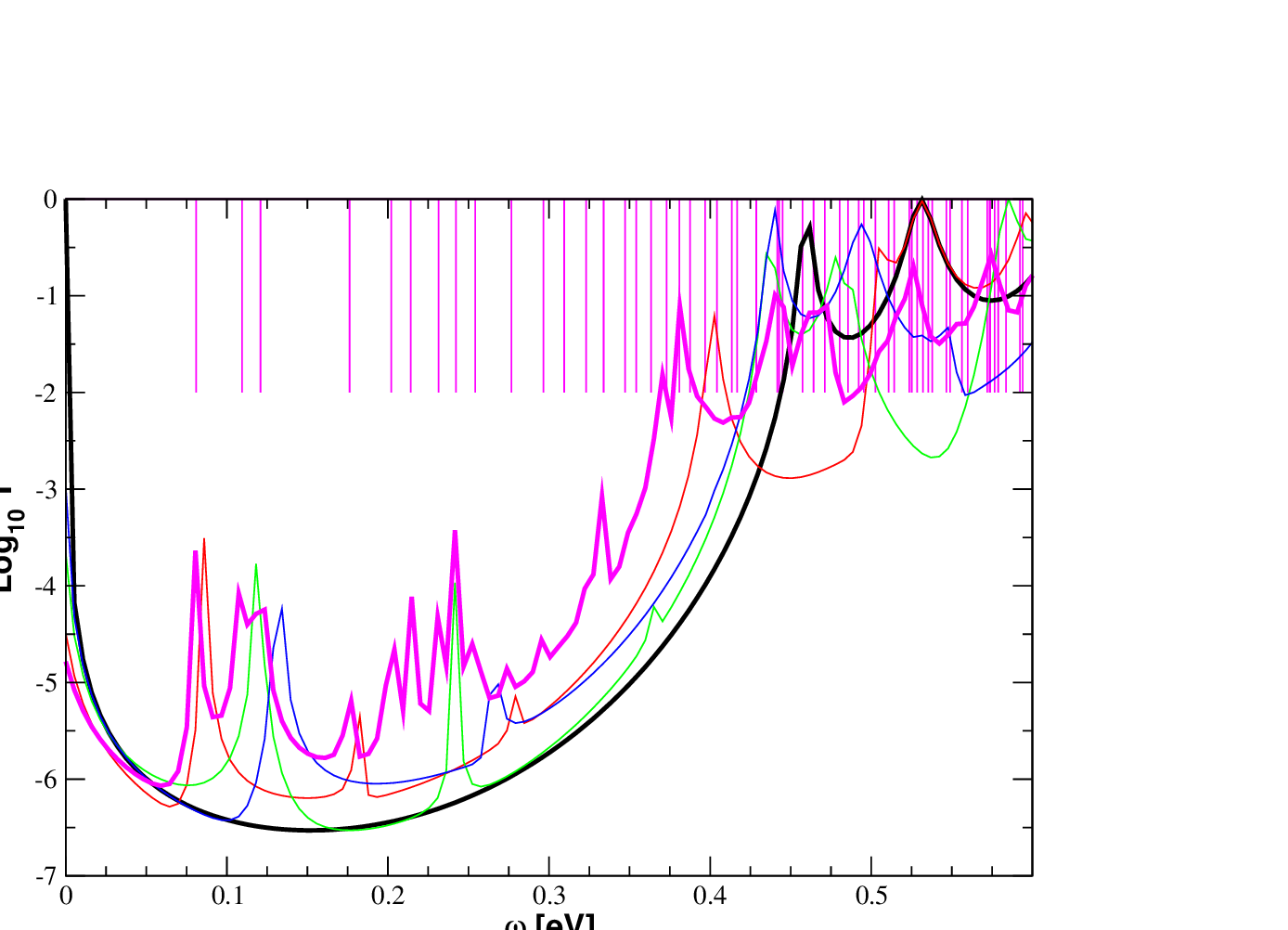}
\end{center}
\caption{Total effective transmission $T_{\rm eff}$ versus 
energy $\omega$ for a molecular wire of length $N=155$. 
The transmission (magenta line) is calculated by including the 
coupling to the three optic modes shown in Fig.\ref{fig1}.
New spectroscopic features appear in the transmission.
The vertical bars give the eigenvalues of the isolated molecular 
Hamiltonian with many levels coupled to the three optic modes.
The transmission in the absence of $e$-ph coupling is
also shown (black line) as well as the transmission obtained for a
coupling to a single mode (thin solid line) as in Fig. \ref{fig1}.}
\label{fig2}
\end{figure}

However for realistic molecules, the $e$-ph coupling matrix is 
non-diagonal and at the same time the injected charge may couple 
simultaneously to different vibration modes.
Calculations including simultaneous coupling to the three
optic modes mentioned above were performed.
New spectroscopic information is obtained, as shown in Fig.\ \ref{fig2}.
The transmission is not simply a superposition of the contribution 
of the three different modes shown in Fig.\ \ref{fig1}; instead, 
new peaks appear.
These features can be interpreted in terms of quantum interferences
between the different elastic and inelastic electron paths as well
as in terms of beating effects. 
All these features in the transmission correspond to the eigenvalues 
of the coupled $e$-ph molecular Hamiltonian $H$, provided coupling 
of the many molecular levels to the three different vibration modes 
is included.
A detailed analysis of the origins of such features is beyond the 
scope of the present paper; however, these effects show that the 
general form for the $e$-ph coupling Eq.(\ref{Heph}) in realistic 
systems creates features in the transmission that cannot be obtained
in terms of single vibration mode analysis.

This is a very important result when one considers the interpretation 
of recent IETS experiments in molecular junctions\footnote{Although the 
total effective transmission is not the differential conductance of 
the IETS spectra, it contains the same spectroscopic 
informations of the coupled $e$-ph system.}.
Our results shown that special care is needed when assigning features
in the transmission or the current to specific vibration modes of the
molecular junction. Calculations for realistic three-dimensional 
geometries in molecular junctions are currently under investigation.


\begin{thebibliography}{99}

\bibitem{kouwenhoven01}
Kouwenhoven, L.P., Austing, D.G., \& Tarucha, S. (2001){\it Rep.\ Prog.\ Phys.} {\bf 64} 701--736.

\bibitem{emin76}
Emin, D. \& Holstein, T. (1976) {\it Phys.\ Rev.\ Lett.} {\bf 36} 323--326.

\bibitem{WHo02}
Ho, W. (2002) {\it J. Chem. Phys.} {\bf 117}, 11033-11061.

\bibitem{HPark00}
Park, H., Park, J., Lim, A.K.L., Anderson, E.H., Alivisatos, A.P. \&
McEuen, P.L. (2000)
{\it Nature} {\bf 407}, 57-60.

\bibitem{JKushmerick04}
Kushmerick, J.G., Lazorcik, J., Patterson C.H. \& Shashidhar, R.
(2004) {\it Nano Lett.} {\bf 4}, 639-642.

\bibitem{WWang04}
Wang, W., Lee, T., Kretzschmar I. \& Reed, M.A.
(2004) {\it Nano Lett.} {\bf 4}, 643-646.

\bibitem{LHYu04}
Yu, L.H., Keane, Z.K., Ciszek, J.W., Cheng, L., Stewart, M.P.,
Tour, J.M. \& Natelson, D.
(2004) {\it Phys. Rev. Lett.} {\bf 93}, 266802.

\bibitem{AHeeger01}
Heeger, A.J.
(2001) {\it Rev. Mod. Phys.} {\bf 73}, 681-700.

\bibitem{BPersson87}
Persson, B.N.J. \& Baratoff A.
(1987) {\it Phys. Rev. Lett.} {\bf 59}, 339-342.

\bibitem{NLorente00}
Lorente, N. \& Persson, M. 
(2000) {\it Phys. Rev. Lett.} {\bf 85}, 2997-3000.

\bibitem{TTodorov98}
 Todorov, T.N.
(1998) {\it Phil. Mag.} B {\bf 77}, 965-73.

\bibitem{DiVentra04}
Chen, Y.C., Zwolak, M. \& Di Ventra, M.
(2004) {\it Nano Lett.} {\bf 4}, 1709-1712.

\bibitem{JBonca95}
Bon\v{c}a, J. \& Trugman S.A., 
(1995) {\it Phys. Rev. Lett.} {\bf 75}, 2566-2569.

\bibitem{HNess99}
Ness, H. \& Fisher, A.J.
(1999) {\it Phys. Rev. Lett.} {\bf 83}, 452-455.

\bibitem{KHaule99}
Haule, K. \& Bon\v{c}a, J. 
(1999) {\it Phys. Rev. B} {\bf 59}, 13087-13093.

\bibitem{NMingo00}
Mingo, N. \& Makoshi K.
(2000) {\it Phys. Rev. Lett.} {\bf 84}, 3694-3697.
 
\bibitem{HNess01}
Ness, H., Shevlin, S.A. \& Fisher, A.J.
(1999) {\it Phys. Rev. B} {\bf 63}, 125422.

\bibitem{HNess02}
Ness, H. \& Fisher, A.J.
(2002){\it Europhys. Lett.} {\bf 57}, 885-891.

\bibitem{ATroisi03}
Troisi, A., Ratner, M.A. \& Nitzan, A. 
(2003) {\it J. Chem. Phys.} {\bf 118}, 6072-6082.

\bibitem{PHanggi04}
Petrov, E.G., May, V. \& H\"anggi, P.
(2004) {\it Chem. Phys.} {\bf 296}, 251-266.

\bibitem{ANitzan01}
Nitzan, A.
(2001) {Annu. Rev. Phys. Chem.} {\bf 52}, 681-750.

\bibitem{CJoachim02}
Hliwa, M. \& Joachim, C.
(2002) {\it Phys. Rev. B} {\bf 65}, 085406.

\bibitem{TMii03}
Mii, T., Tikhodeev, S.G. \& Ueba, H.
(2003) {\it Phys. Rev. B} {\bf 68}, 205406.

\bibitem{MGalperin04}
Galperin, M., Ratner, M.A. \& Nitzan, A.
(2004) {\it Nano Lett.} {\bf 4}, 1605-1611.

\bibitem{APecchia04}
Pecchia, A., Di Carlo, A., Gagliardi, A., Sanna, S., Frauenhein,
T. \& Gutierrez, R.
(2004) {\it Nano Lett.} {\bf 4}, 2109-2114.

\bibitem{YAsai04}
Asai, Y.
(2004) {\it Phys. Rev. Lett.} {\bf 93}, 246102.
 
\bibitem{TFrederiksen04}
Frederiksen, T., Brandbyge, M., Lorente, N. \& Jauho, A.P.
(2004) {\it Phys. Rev. Lett.} {\bf 93}, 256601.
 
\bibitem{PCornaglia04}
Cornaglia, P.S., Ness, H. \& Grempel, D.R.
(2004) {\it Phys. Rev. Lett.} {\bf 93}, 147201.

\bibitem{AHorsfield04}
Horsfield, A.P., Bowler, D.R., Fisher, A.J., Todorov, T.N. \& Montgomery M.J.
(2004) {\it J. Phys.: Condens. Matter} {\bf 16}, 3609-3622.

\bibitem{Glauber69}
Cahill, K.E. \& Glauber, R.J. (1969) {\it Phys.\ Rev.} {\bf 177}, 1857-1881. 

\bibitem{NWingreen89}
Wingreen, N.S., Jacobsen, K.W. \& Wilkins, J.W.
(1989) {\it Phys. Rev. B} {\bf 40}, 11834-11850.

\bibitem{YMeir92}
Meir, Y. \& Wingreen, N.S.
(1992) {\it Phys. Rev. Lett.} {\bf 68}, 2512-2515.

\bibitem{Flensberg03}
Flensberg, K. (2003) {\it Phys.\ Rev. B} {\bf 68}, 205323.

\bibitem{JReichert03}
Reichert, J., Weber, H.B., Mayor, M. \& L\"ohneysen H.V.
(2003) {\it Appl. Phys. Lett.} {\bf 82}, 4137-4139.

\end{thebibliography}
\end{document}